\documentclass[10pt,twocolumn,letterpaper]{article}

\usepackage[cvprfinal]{cvpr}  

\usepackage{amssymb}
\makeatletter

\newcommand{\Rmnum}[1]{\expandafter\@slowromancap\romannumeral #1@}
\makeatother

\definecolor{cvprblue}{rgb}{0.21,0.49,0.74}
\usepackage[pagebackref,breaklinks,colorlinks,citecolor=cvprblue]{hyperref}

\def\paperID{2288} 
\def\confName{CVPR}
\def\confYear{2025}

\title{Prosody-Enhanced Acoustic Pre-training and Acoustic-Disentangled \\ Prosody Adapting for Movie Dubbing} 

\author{Zhedong Zhang$^{1,2}$\thanks{This work is done during the intern in VIPL group, ICT, CAS.}\quad Liang Li$^2$\thanks{Corresponding author.}\quad Chenggang Yan$^1$\quad Chunshan Liu$^1$ \\ \quad Anton van den Hengel$^3$ \quad Yuankai Qi$^4$\\
$^1$Hangzhou Dianzi University\quad $^2$Institute of Computing Technology, Chinese Academy of Sciences \\ $^3$ University of Adelaide\quad $^4$ Macquarie University}
\begin{document}
\maketitle
\begin{abstract}

Movie dubbing describes the process of transforming a script into speech that aligns temporally and emotionally with a given movie clip while exemplifying the speaker’s voice demonstrated in a short reference audio clip.
This task demands the model bridge character performances and complicated prosody structures to build a high-quality video-synchronized dubbing track.
The limited scale of movie dubbing datasets, along with the background noise inherent in audio data, hinder the acoustic modeling performance of trained models.
To address these issues, we propose an acoustic-prosody disentangled two-stage method to achieve high-quality dubbing generation with precise prosody alignment.
First, we propose a prosody-enhanced acoustic pre-training to develop robust acoustic modeling capabilities.
Then, we freeze the pre-trained acoustic system and design an acoustic-disentangled framework to model prosodic text features and dubbing style while maintaining acoustic quality.
Additionally, we incorporate an in-domain emotion analysis module to reduce the impact of visual domain shifts across different movies, thereby enhancing emotion-prosody alignment.
Extensive experiments show that our method performs favorably against the state-of-the-art models on two primary benchmarks.
The demos and source code are available at \textcolor{blue}{\href{https://zzdoog.github.io/ProDubber/}{here}}.
\end{abstract}

\section{Introduction}
\label{Intro}

\begin{figure}[t]
    \centering
    \includegraphics[width=1\linewidth]{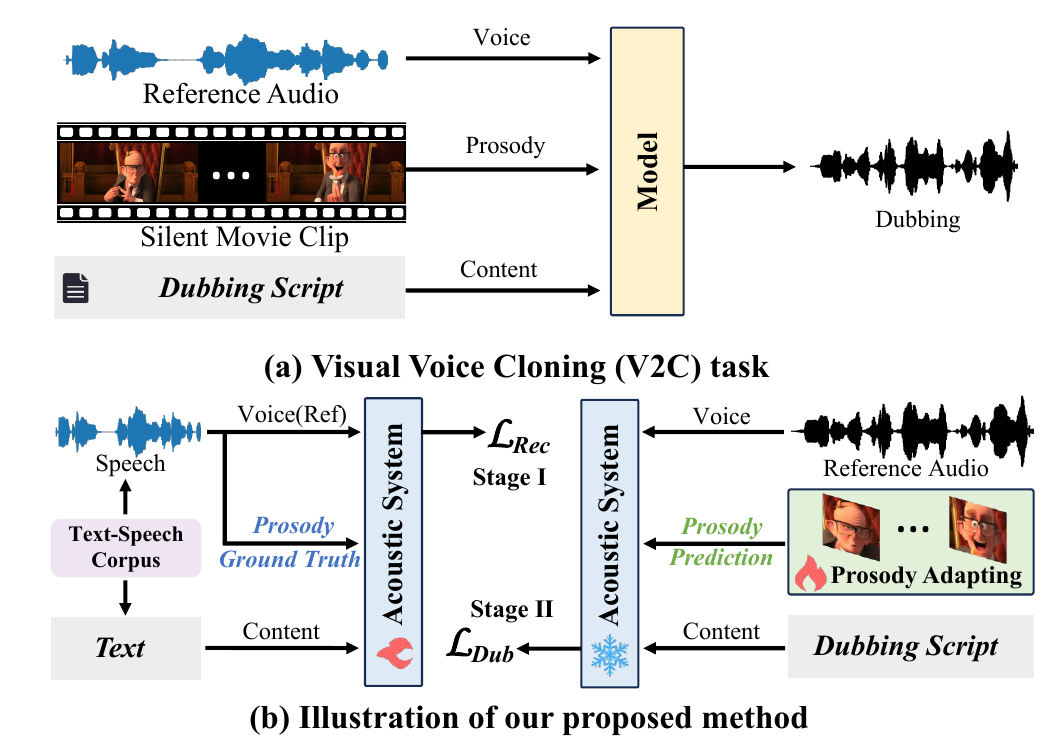}
    \vspace{-0.4cm}
    \caption{(a) Illustration of the V2C task. 
             (b) Illustration of the proposed acoustic pre-training stage (Stage \Rmnum{1}) and prosody adapting stage (Stage \Rmnum{2}), which aim to generate dubbing with high acoustic quality and aligned prosody.
             }
    \label{Fig-intro}
        \vspace{-0.4cm}
\end{figure}

Movie Dubbing, also known as Visual Voice Cloning (V2C)~\cite{V2C}, aims to generate speech from scripts using a specified voice conditioned by a single short reference audio while ensuring prosody alignment with the character’s performance in terms of both pronunciation duration and intonation (as shown in Figure~\ref{Fig-intro} (a)).
It promises significant potential in real-world applications such as film post-production, media production, and personal speech AIGC.


Compared to traditional speech synthesis~\cite{fs1, fs2, stylespeech, StyleTTS2, adaspeech} or Voice Cloning~\cite{Nerual_VC} tasks, movie dubbing presents greater challenges.
It demands the ability not only to align the prosody with complex character performances but also to maintain high acoustic quality even when managing diverse and exaggerated prosody inherent in dubbing.





Previous works achieve significant progress in aligning prosody with the performances of movie characters.
V2C-Net~\cite{V2C} incorporates visual features from each frame of the movie clip into the mel-spectrogram frame-level prosody modeling.
HPMDubbing~\cite{HPMDubbing} employs a hierarchical design, where lip, face, and video-level features correspond to pronunciation duration, prosody, and the overall video atmosphere, respectively.
StyleDubber~\cite{styledubber} uses a fine-grained adaptor to bridge the speaker's style and character's emotion states at the phoneme level, facilitating joint modeling of prosody.
However, existing methods overlook the visual domain discrepancies arising from variations in movie styles and other visual elements, such as color grading, which impact the stability of prosody alignment.

The nature of film acting means movie dubbing inherently involves more exaggerated and diverse prosody variations compared to traditional speech, yet the difficulty of its acoustic modeling is often underestimated.
In addition, certain characteristics of movie dubbing datasets, such as their limited scale (primarily due to copyright restrictions) and the suboptimal audio quality caused by unavoidable movie background noise~\cite{spk2dub}, further hinder methods that solely rely on dubbing datasets for training to achieve satisfied acoustic quality~\cite{V2C, facetts, HPMDubbing, styledubber}.
Speaker2Dubber~\cite{spk2dub} attempts to enhance the model's acoustic modeling capability by pre-training the phoneme encoder on a large-scale text-to-speech corpus.
This two-stage training strategy improves the model's pronunciation accuracy, 
however, the pre-trained phoneme encoder alone cannot eliminate the impact of low audio quality in dubbing datasets, or modeling complex prosody variation, leading to a significant degradation in final acoustic quality.

To address these issues, we propose a dubbing method that incorporates a prosody-enhanced acoustic pre-training stage and an acoustic-disentangled prosody adapting stage to achieve high-quality dubbing with well-aligned prosody (as shown in Figure~\ref{Fig-intro} (b)).
Specifically, in the prosody-enhanced acoustic pre-training stage, we pre-train the entire acoustic system of the model using a text-speech corpus with enhanced prosody.
This ensures the acoustic modeling capability in audio quality and managing the complex dubbing prosody variations.
%
We thus freeze the acoustic system in the second stage and design a disentangled framework to model prosodic features from the dubbing script and audio by introducing a prosodic text BERT encoder and a prosodic style diffusion module.
These disentangled prosodic features are further bridged with the in-domain emotion analysis output feature, aligning the prosody with the character's emotion.
Lip motion features are then utilized to predict pronunciation duration and upsample the text content to match the mel-spectrogram length.
Finally, the pre-trained audio decoder takes the upsampled text content and predicted prosody as input, integrating the speaker voice via AdaIN layer~\cite{AdaIN}, and generates dubbing with high acoustic quality and aligned prosody.

The main contributions of the paper are as follows:
\begin{itemize}
\item We propose a two-stage dubbing method with a prosody-enhanced acoustic pre-training to enhance the model’s ability in high-quality acoustic modeling and handling complex prosodic variations.
\item To maintain the acoustic quality in the prosody adapting stage, we freeze the pre-trained acoustic system and devise an acoustic-disentangled prosody adapting framework by introducing prosodic modeling modules corresponding to both the script and the required dubbing style.
\item We design an in-domain emotion analysis module in the prosody adapting stage to reduce the instability caused by visual domain discrepancies across different movies.
\item Extensive experiments demonstrate the proposed method achieves favorable performance compared to several state-of-the-art approaches on two major benchmarks, validating the effectiveness of our approach.
\end{itemize}
\begin{figure*}[t]
 \centering
  \resizebox{\linewidth}{!}{\includegraphics{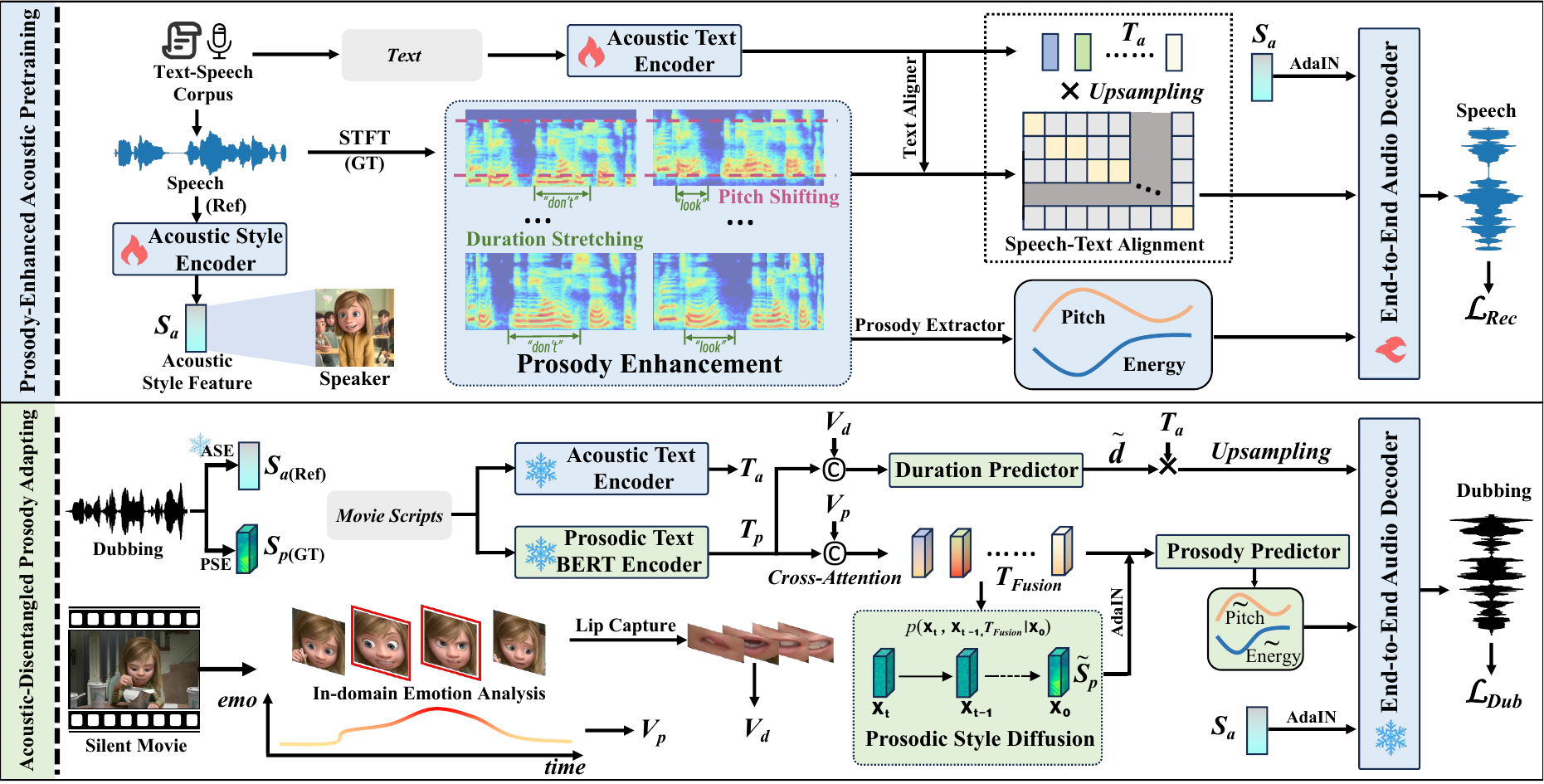}}
  \caption{
  The main architecture of the proposed method.
  In the Prosody-Enhanced Acoustic Pre-training stage (Section~\ref{PEAP}), we pre-train the acoustic system using a prosody-enhanced text-speech corpus.
  In the Acoustic-Disentangled Prosody Adapting stage (Section~\ref{DPA}), we freeze the acoustic system and employ a disentangled framework to bridge the prosody of dubbing with the character's performance using In-Domain Emotion Analysis (Section~\ref{IDEA}), thus generating dubbing with aligned prosody and maintain high acoustic quality.}
  \label{method}
\vspace{-0.4cm}
\end{figure*}
\section{Related Work}
\subsection{Speech Synthesis}
With the rapid development of deep learning~\cite{tian2023transformer, zhang2024deep, huijuan2023improved, peng2024fast,cui2024stochastic, yin2023reducing, cong2024emodubber, deng2024multi, gait3d_v1,tu2023self, tu2024context, tu2024distractors, tu2024smart, liu2022entity, ye2022unsupervised}, the speech synthesis task has advanced significantly in recent years.
From the autoregressive generation architecture of models like WaveNet~\cite{wavenet} to the non-autoregressive parallel architecture of the FastSpeech series~\cite{fs1,fs2}, both the quality and speed of speech generation have improved substantially.
Recently, the NaturalSpeech~\cite{naturalspeech, naturalspeech2, naturalspeech3} and StyleTTS~\cite{styletts,StyleTTS2} series have utilized factorized codec and diffusion models, as well as adversarial training with large speech language models, respectively, to achieve high-quality speech synthesis that more closely approaches human-level performance.
Additionally, many other works based on other generative models~\cite{prodiff, gradtts, speakreadrompt, audiogpt, megatts, zeroshotspeech} such as flow matching~\cite{glowtts, voiceflow, li2024flowgananomaly} are also advancing the progress in speech synthesis.
Despite the progress has been made, they cannot be directly applied to movie dubbing tasks because they lack the design of modeling prosody from input movie clips.

\subsection{Visual Voice Cloning}
The V2C task~\cite{V2C} requires the model to align the prosody of the generated speech with the input video to achieve automatic movie dubbing. 
To this end, HPMDubbing~\cite{HPMDubbing} proposes a hierarchical prosody learning structure, aligning the dubbing duration, prosody, and overall tone with the lip movements, facial expressions, and overall context of the video. 
StyleDubber~\cite{styledubber}, on the other hand, designs a fine-grained prosody matching module for character voice, incorporating phoneme-level features and emotional variations to model the dubbing style of the characters.
Subsequently, Speaker2Dubber~\cite{spk2dub} addresses the difficulty of learning accurate pronunciation through training solely on dubbing datasets by pre-training its phoneme encoder on text-speech corpus.
Despite these advancements, current dubbing models are still significantly influenced by the quality of the dubbing dataset and struggle to achieve high acoustic quality.


\section{Method}
\subsection{Overview}
The target of the overall movie dubbing task is:
\begin{equation}
     \tilde{A} _{Dub} = \mathrm{Model}(A_{Ref}, T_{d}, V_{m}),
\end{equation}
where the $\tilde{A} _{Dub}$ is the generated dubbing and $A_{Ref}, T_{d}, V_{m}$ are the reference audio, dubbing scripts, and the input silent movie clip respectively.

The main architecture of the proposed two-stage method is shown in Figure~\ref{method}.
In the prosody-enhanced acoustic pre-training stage, we first design a prosody enhancement strategy to address the prosody domain differences between movie dubbing and traditional speech and apply it to a large-scale, high-quality text-speech corpus.
Then, we use the enhanced corpus to pre-train the acoustic style encoder (ASE), acoustic text encoder, and end-to-end audio decoder of the dubbing model, enabling the acoustic system to accurately generate high-quality speech with corresponding content and speaker voice.
Next, in the acoustic-disentangled prosody adapting stage, we utilize an acoustic-prosodic disentangled framework by introducing the prosodic style diffusion and prosodic text BERT encoder.
This framework adapts the prosody while keeping the acoustic system frozen, thus maintaining the acoustic quality of generated dubbing.
We also propose In-Domain Emotion Analysis to better bridge the dubbing prosody with the character's emotion, which improves the audiovisual consistency.
We detail each module below.

\subsection{Prosody-Enhanced Acoustic Pre-training}
\label{PEAP}
Acoustic quality forms the foundation of movie dubbing.
The goal of prosody-enhanced acoustic pre-training is to provide a solid basis for high-quality acoustic modeling for movie dubbing through the pre-training of the acoustic system in the dubbing model. 
Below, we detail this pre-training approach in two aspects: the pre-training objectives and the prosody enhancement strategy.

\subsubsection{Pre-training Objectives}
Unlike the target of the movie dubbing task, the input of this stage does not contain the movie clip:
\begin{equation}
     \tilde{A} _{speech} = \mathrm{Model}(A_{Ref}, T_{s}),
\end{equation}
where the $T_{s}$ is the script from the text-speech corpus.
Nevertheless, the acoustic information ultimately used to generate speech audio comprises three components which are the same as the movie dubbing: content information derived from the text, speaker voice obtained from reference audio, and the prosody of the speech.
The only difference lies in whether the prosody and duration of each phoneme need to align with the given video. 
Since the text-speech corpus used in this stage does not include the corresponding video clip, we directly utilize the ground truth duration and prosody during this stage.

We adopt an architecture similar to StyleTTS~\cite{styletts} for acoustic pre-training. 
First, the text in the corpus is converted from a grapheme sequence to a phoneme sequence using a G2P module. 
We then use a bidirectional LSTM network to extract acoustic text features from the phoneme sequence. 
Next, an acoustic style encoder extracts timbre features from either the ground truth or other speech samples from the same speaker.
A prosody extractor is employed to extract mel-spectrogram frame-level prosody (pitch and energy curve) from the mel-spectrogram of the ground truth speech.
To upsample the acoustic text sequence to match the length of the mel-spectrogram, we use a monotonic aligner pre-trained on the LibriSpeech dataset for the automatic speech recognition (ASR) task to obtain the ground truth duration of each phoneme:
\begin{equation}
\begin{split}
 &T_{pho} = \mathrm{G2P}(T_s)\in \mathbb{R}^{L_{pho}}, 
\\&T_a = \mathrm{Acoustic Text Encoder}(T_{pho})\in \mathbb{R}^{L_{pho}\times d_{m}},
\\&S_{a} = \mathrm{Acoustic Style Encoder}(A_{Ref})\in \mathbb{R}^{d_m},
\\&Align = \mathrm{TextAligner}(Mel_{s}, T_{Pho})\in \mathbb{R}^{L_{pho}\times L_{mel}},
\\&p,n = \mathrm{ProsodyExtractor}(Mel_s)\in \mathbb{R}^{L_{mel}},
\end{split}
\end{equation}
where $T_a$ and $S_a$ are the acoustic text feature and acoustic style feature, respectively.
$L_{pho}$ and $L_{mel}$ denote the length of the phoneme sequence and desired mel-spectrogram, respectively.
$d_m$ denotes the hidden size of the model.
$Align$ is the monotonic attention map between the phoneme sequence and the mel-spectrogram of ground truth speech $Mel_s$.
$p, n$ is the ground truth pitch and energy curve.

Then, a HiFi-GAN~\cite{hifigan} based end-to-end audio decoder takes the extracted prosody and the upsampled acoustic text features as inputs, while using Adaptive Instance Normalization (AdaIN)~\cite{AdaIN} to integrate the acoustic style information and reconstruct the original ground truth waveform:
\begin{equation}
\begin{split}
&\mathrm{AdaIN}(c, S_a) = P_\sigma (S_a)\frac{c-\mu (c)}{\sigma (c)}+ P_\mu (S_a), \\
&\tilde{A} _{speech} = \mathrm{AudioDecoder}(T_a \cdot Align, p, n, S_a), 
\end{split}
\end{equation}
where $c$ is the single channel of the feature maps, $\mu(\cdot)$ and $\sigma(\cdot)$ denote the channel mean and standard deviation.
$P_\mu$ and $P_\sigma$ are the linear projections for predicting the adaptive gain and bias using the acoustic style feature $S_a$.

The training objective at this stage is a reconstruction loss for speech:
\begin{equation}
    \mathcal{L}_{Rec} = \left \| A_{speech} - \tilde{A}_{speech} \right \| _1.
\end{equation}

\subsubsection{Prosody Enhancement Strategy}
By pre-training the acoustic system using a text-speech corpus as mentioned above, the dubbing model gains the ability to convert text into high-quality speech. 
However, the speech in traditional text-to-speech corpora typically comes from recordings of audiobooks or blogs, which often exhibit flat and monotonous prosody. In contrast, the prosody of movie dubbing tends to vary significantly depending on the plot or performance, particularly in terms of pitch variation and the duration of certain lines.


We design two prosody enhancement strategies to address the prosodic domain discrepancy between movie dubbing and traditional speech: pitch shifting and duration stretching. 
Pitch shifting alters the overall pitch of the speech by shifting it to higher or lower frequency bands, simulating the pitch variations found in dubbing.
Duration stretching scales the length of the speech to simulate the varied speech pace observed in movie dubbing.

During the implementation, we find that directly manipulating pitch or duration in the time domain significantly degrades speech quality and reduces its naturalness. 
Therefore, we choose to convert the speech into mel-spectrograms and apply the enhancements in the frequency domain. 
We then use a vocoder~\cite{hifigan} to convert the modified speech back to the time domain for end-to-end pre-training.
Specifically, for the pitch shifting operation, we shift the entire spectrogram up or down by a specified number of frequency bands and fill the excess frequency bands with -80db to maintain the shape of spectrogram.
For duration stretching, we use linear interpolation and downsampling operations to expand or shorten the length of spectrograms.

\subsection{In-Domain Emotion Analysis}
\label{IDEA}
To obtain fine-grained character emotion features at the video frame level, we first extract the face region of each frame of movie clip $V_{face}$ via $\mathrm{S^3FD}$ face detection~\cite{s3fd}, then utilize an emotion face-alignment network ($\mathrm{EmoFAN}$~\cite{emofan}) to encode the face region to emotion features $V_{emo}$ following~\cite{HPMDubbing}:
\begin{equation}
    V_{emo}=\mathrm{EmoFAN}(\mathrm{S^3FD}(V_{m})) \in \mathbb{R}^{L_v\times d_{m}},
\end{equation}
where the $L_v$ denotes the frame number of the video.

To ensure that the fine-grained video frame-level emotion features more accurately reflect emotional changes rather than visual characteristics of the movie, we first normalize them:
\begin{equation}
V_{emo}' = \Sigma^{-\frac{1}{2}} (V_{emo} - \mu)\in \mathbb{R}^{L_v\times d_{m}},
\end{equation}
where  $V_{emo}'$ denotes the normalized feature, $\mu$ is the mean of the feature sequence, and $\Sigma$ is the covariance matrix of sequence. 
The term \( \Sigma^{-\frac{1}{2}} \) represents the inverse square root of the covariance matrix, which ensures that the normalized features have zero mean and unit covariance.

Simultaneously, to reflect the absolute emotional state of characters in the video, we use a video-level atmosphere feature~\cite{HPMDubbing} to adjust the normalized emotion features, thereby capturing the character's emotional state and changes:
\begin{equation}
\begin{split}
    V_p =  \alpha \cdot V_{emo}' &+ \beta \in \mathbb{R}^{L_v\times d_{m}},\\
    \alpha = f_{\alpha}(V_{ato}), &\quad \beta = f_{\beta}(V_{ato}),
\end{split}
\end{equation}
where $V_{ato}\in \mathbb{R}^{d_{m}}$ is the video-level atmosphere feature, $f_{\alpha}$ and $f_{\beta}$ are learnable linear projections.
$V_p$ denotes the video-frame level emotion prosodic features.


\subsection{Acoustic-Disentangled Prosody Adapting}
\label{DPA}
To ensure that the model does not compromise acoustic quality during training on the dubbing dataset, we freeze the entire pre-trained acoustic system at this stage.
Unlike traditional TTS tasks that only require modeling prosody from the input text and the voice of reference audio, movie dubbing demands a more complex prosody modeling that bridges the character's facial expressions from visual modality.
Relying solely on the frozen acoustic text encoder and acoustic style encoder is insufficient to handle the complex prosody alignment.

To this end, we propose an acoustic-disentangled framework by introducing a prosodic text BERT encoder (PTBE)~\cite{pl=bert} and a prosodic style encoder (PSE) to extract prosodic features from the script text and non-timbre prosodic features from ground truth dubbing, respectively:
\vspace{-0.6cm}
\begin{equation}
\begin{split}
\\&T_p = \mathrm{PTBE}(T_{pho})\in \mathbb{R}^{L_{pho}\times d_{m}},
\\&S_{p} = \mathrm{ProsodicStyleEncoder}(A_{Dub})\in \mathbb{R}^{d_m}. \\
\end{split}
\end{equation}
After extracting prosodic features from both the text and the dubbing audio, we employ a multi-head cross-attention module to connect phoneme-level text prosodic features with emotion features, enabling the modeling of prosody at the phoneme level based on character emotion variations:
\vspace{-0.2cm}
\begin{equation}
\begin{split}
T^{k}_{Fusion} &= softmax(\frac{Q_{pho}^\top K_{emo}}{\sqrt{d_{m}}})V_{emo} \in \mathbb{R}^{L_{pho}\times \frac{d_{m}}{n\_head} }, \\
Q_{pho} &= W^Q_jT_{p}^\top, K_{emo} = W^{K_p}_jV_{p}^\top, V_{p} = W^{V_p}_jV_{p}^\top,\\
\end{split}
\end{equation}
where $k$ denotes $k$-th head's output, $W_{j}^*$ are learnable parameter matrix, and $T_{Fusion}$ is phoneme-level prosodic feature with emotion variation information.

To model the prosodic style feature of dubbing, we design a prosodic dtyle diffusion module that takes $T_{Fusion}$ as input and samples the prosodic style feature from noise:
\begin{equation}
\begin{split}
p_\theta(S_{t-1} | S_t) &= \mathcal{N}(S_{t-1}; \mu_\theta(S_t, t), \Sigma_\theta(S_t, t)),\\
\tilde{S}_p &= p_\theta(S_0 | T_{Fusion}),
\end{split}
\end{equation}
where the $p_\theta$ is the learnable denoise distribution and the $S_{t-1}$ is the feature of previous time step. Specifically, in the first half of the training process, we train the prosodic style encoder to better assist the prosody predictor in prosody prediction. 
In the second half of the training, we introduce the prosodic style diffusion module. 
This approach ensures more stable convergence during training.

Subsequently, we employ the AdaIN module to integrate the predicted prosodic style feature into the $T_{Fusion}$, which is then input into the prosody predictor for specific prosody prediction:
\begin{equation}
\tilde{p}, \tilde{n} = \mathrm{ProsodyPredictor}(\mathrm{AdaIN}(T_{Fusion}, \tilde{S}_p)),
\end{equation}
where the $\tilde{p}, \tilde{n}\in \mathbb{R}^{L_{pho}}$ are the prediction of prosody attributes (pitch and energy) based on the prosodic features and the visual emotion feature.

In terms of aligning the duration of pronunciation with lip movements, since this alignment is solely related to the prosody of pronunciation and lip movements,
which is independent of character emotions and both kinds of style, the input to the duration predictor is the cross-attention fusion of the prosodic text feature and the lip motion feature.
Then, we upsample the acoustic text features to the spectrogram length based on the alignment built from the duration prediction, and the audio decoder takes the upsampled acoustic text feature, the acoustic style feature, and the prosody prediction as input, generating the dubbing audio:
\begin{equation}
\begin{split}
&\tilde{d}=\mathrm{DurationPredictor}(\mathrm{CA}(T_p, V_d, V_d))\in \mathbb{R}^{L_{pho}},\\
&\title{\tilde{A}_{Dub}}=\mathrm{AudioDecoder}(T_a\cdot\tilde{Align},\tilde{p}, \tilde{n}, S_a),
\\
\end{split}
\end{equation}
where the $V_d\in \mathbb{R}^{L_{v}\times d_{m}}$ is the lip motion feature extracted from the lip region of movie clips following~\cite{HPMDubbing}.
$\mathrm{CA}(\cdot)$ denotes the cross-attention, we use $T_p$ as query and $V_d$ as key and value here to get the phoneme-level duration features.
The duration prediction is scaled according to the length of the movie clip to ensure the alignment between total video duration and total dubbing duration.
It is noteworthy that $\tilde{p}$ and $\tilde{n}$ are also upsampled before being input into the decoder to match the spectrogram length.

\begin{table*}[!t]
  \centering
    \caption{
  Results on V2C-Animation benchmark. 
  For the Dub 1.0 setting, we use the ground truth audio as reference audio,
  for the Dub 2.0 setting, we use the non-ground truth audio from the same speaker within the dataset as the reference audio which is more aligned with practical usage in dubbing.
  The method with “*” refers to a variant taking video embedding as an additional input following~\cite{V2C}.
  “T-S” indicates whether this method employs a two-stage training approach same as~\cite{spk2dub}.
  The same setup is applied to the GRID benchmark.
  }
  \vspace{-0.2cm}
  \resizebox{1.0\linewidth}{!}
  {

    \begin{tabular}{c|c|cccccc|ccc}
    \hline
    Setting & & \multicolumn{6}{c|}{Dub 1.0} & \multicolumn{3}{c}{Dub 2.0} \\ 
    \toprule
    Methods & \small{T-S}
    & SECS (\%) $\uparrow$ 
    & WER (\%) $\downarrow$
    & UT-MOS $\uparrow$
    & EMO-ACC (\%) $\uparrow$   
    & MCD-DTW $\downarrow$ 
    & MCD-DTW-SL $\downarrow$
    & SECS (\%) $\uparrow$ 
    & WER (\%) $\downarrow$
    & UT-MOS $\uparrow$   \\
    \midrule
    GT & - & 100.00 & 25.55  & 2.26 & 99.96 & 0.00 & 0.00 & 100.00 & 22.55 & 2.26\\
    GT Mel + Vocoder & - & 96.96 & 24.40 & 2.18& 97.09  & 3.77 & 3.80 & 96.96 & 24.40 & 2.18\\
    \midrule
    StyleSpeech*~\cite{stylespeech} & \text{\sffamily X} & 42.53 & 108.00 & 1.41 & 42.53   & 11.62 & 14.23 &  75.67 & 82.48 & 1.81\\
    Zero-shot TTS*~\cite{zeroshottts} & \text{\sffamily X} & 48.93  & 68.05 & 1.72 & 43.97 & 10.03  &  12.01 & 47.55 & 58.81 & 1.63\\
    V2C-Net~\cite{V2C} & \text{\sffamily X} & 40.61 & 73.08 & 1.32 & 43.08 & 14.12 & 18.49 & 34.07  & 61.61 & 1.30 \\
    HPMDubbing~\cite{HPMDubbing} & \text{\sffamily X} & 53.76 & 164.16 & 1.31 & 46.61 & 11.12  & 11.22 & 31.42 & 171.03 & 1.30 \\
    Face-TTS~\cite{facetts} & \text{\sffamily X} & 52.81  & 201.13 & 1.31 & 44.04 &  13.44 & 26.94 & 51.98 & 200.18 & 1.31\\
    StyleDubber~\cite{styledubber} & \text{\sffamily X} & \textbf{82.26} & 31.49 & 1.89 & 45.62 & 9.37 & 9.46 & \textbf{81.27} & 31.70 & 1.95\\
    \midrule
    StyleSpeech*~\cite{stylespeech} & \checkmark & 58.94 & 38.35 & 2.44 & 46.27 & 10.10 &  13.12 & 55.90 & 63.97 & 1.89 \\
    V2C-Net~\cite{V2C} & \checkmark & 69.00 & 23.02 & 2.35 & 46.15 & 9.97 & 10.97 & 69.30 & 23.15 & 2.32 \\
    Speaker2Dubber~\cite{spk2dub} & \checkmark & 81.50 & 17.51 & 2.41 & 46.80 & 9.46 & 9.65 & 79.86 & 17.33 & 2.42\\
    \midrule
    Ours & \checkmark & 75.46 & \textbf{8.04} & \textbf{3.10} & \textbf{48.93} & \textbf{9.29} & \textbf{9.32} & 75.39 & \textbf{11.50} & \textbf{3.09}\\
    
    \bottomrule
    \end{tabular}
    }

  \label{V2C_results}%
  \vspace{-0.2cm}
\end{table*}%
\subsection{Training}
In the acoustic-disentangled prosody adapting stage, since we freeze all acoustic modules, the training objective for this stage is the sum of the losses for different prosody attributes:
\begin{equation}
\begin{split}
\mathcal{L}_{Dub} = \lambda _1\mathcal{L}_{p}+\lambda _2\mathcal{L}_{n}+\lambda _3\mathcal{L}_{d}+\lambda _4\mathcal{L}_{S_p},
\end{split}
\label{loss}
\end{equation}
where the $\mathcal{L}_{p}$, $\mathcal{L}_{n}$, and $\mathcal{L}_{d}$ are the L1-Loss for pitch, energy and duration respectively.
The $\mathcal{L}_{S_p}$ is the diffusion loss for prosodic style feature modeling.
\section{Experiment}

\subsection{Datasets}
\textbf{V2C-Animation dataset~\cite{V2C}} is a collection of 10,217 video clips from 26 animated movies, featuring aligned text, audio, and video clips from real movies along with speaker emotion annotations.
Our main experimental results are validated on this dataset due to its authenticity and more detailed annotations.


\noindent \textbf{GRID dataset~\cite{GRID}} is a basic benchmark for multi-speaker dubbing. 
It comprises high-quality video recordings of 33 speakers performing 1,000 scripted sentences each, resulting in a total of 33,000 utterances.
All participants are recorded in a noise-free studio with a unified screen background.
V2C-Animation and GRID benchmarks respectively correspond to animated films and basic real-person dubbing, covering a wide range of application scenarios.

\noindent \textbf{LibriTTS dataset~\cite{libritts}} is a substantial collection of speech data designed for training and evaluating multi-speaker text-to-speech (TTS) systems derived from the LibriSpeech~\cite{LibriSpeech}.
It includes multi-speaker recordings, offering a diverse range of voices and speaking styles and meticulously segmented and annotated with text transcripts. 
We employ LibriTTS as the text-speech corpus used in the prosody-enhance acoustic pre-training stage.


\subsection{Implementation Details}
Following~\cite{V2C, HPMDubbing, styledubber, spk2dub}, all the video frames are sampled at 25 FPS. 
During training, we resample all audio file in 24kHz and convert them into mel-spectrograms using FFT size of 2048, hop size of 300, window size of 1200, and frequency bins of 80.
We employ a pre-trained JDC network~\cite{JDC} as the pitch extractor and use the log norm to calculate the energy following~\cite{styletts, StyleTTS2}.
For the text aligner, we adopt the ASR model fine-tuned for the TTS task by~\cite{styletts} to get the ground truth alignment between phoneme and mel-spectrogram.

We use Libri-TTS-460 following~\cite{StyleTTS2} for the acoustic pre-training and apply a 3\% prosody enhancement ratio to trade off the audio quality and prosody variations.
We use the official pre-trained HiFi-GAN~\cite{hifigan} as the vocoder for the prosody enhancement.
We train the acoustic system with 20 epochs in the pre-training stage.
For the prosody adapting stage, we train the model with 30 epochs on the V2C-Animation dataset and 20 epochs for GRID dataset. 
An Adam~\cite{adam} with $\beta_1=0.9$, $\beta_2=0.98$, $\epsilon=10^{-9}$ is used as the optimizer in both the training stages.
The learning rate is set to 0.00625.
The weight in Equation ~\ref{loss} are set to $\lambda_1 = 1$, $\lambda_2 = 1$, $\lambda_3 = 1$, $\lambda_4 = 0.2$.
For a fair comparison, all comparison models are re-trained on the same dataset. 

\begin{table*}[!t]
  \centering
    \caption{
  Results on GRID benchmark. 
  }
  \vspace{-0.2cm}
  \resizebox{1.0\linewidth}{!}
  {

    \begin{tabular}{c|c|ccccc|ccc}
    \hline
    Setting & & \multicolumn{5}{c|}{Dub 1.0} & \multicolumn{3}{c}{Dub 2.0} \\ 
    \toprule
    Methods & \small{T-S}
    & SECS (\%) $\uparrow$ 
    & WER (\%) $\downarrow$
    & UT-MOS $\uparrow$
    & MCD-DTW $\downarrow$ 
    & MCD-DTW-SL $\downarrow$
    & SECS (\%) $\uparrow$ 
    & WER (\%) $\downarrow$
    & UT-MOS $\uparrow$   \\
    \midrule
    GT & - & 100.00 & 22.41 & 3.94 & 0.00 & 0.00 & 100.00 & 22.55 & 3.94\\
    GT Mel + Vocoder & - & 97.57 & 21.41 & 3.57 & 4.10 & 4.15 & 97.57 & 21.41 & 3.57\\
    \midrule
    StyleSpeech*~\cite{stylespeech} & \text{\sffamily X} & 91.06 & 24.83 & 3.71 & 5.87 & 5.98 & 82.57 & 21.42 & 3.57\\
    Zero-shot TTS*~\cite{zeroshottts} & \text{\sffamily X} & 86.54 & 19.13 & 3.60 & 5.71 & 5.99 & 72.91 & 19.35 & 3.62\\
    V2C-Net~\cite{V2C} & \text{\sffamily X} & 80.98 & 47.82 & 2.41 & 6.79 & 7.23 & 75.92 & 49.09 & 2.40\\
    HPMDubbing~\cite{HPMDubbing} & \text{\sffamily X} & 85.11 & 45.51 & 2.14 & 6.49 & 6.78 & 61.32 & 44.15 & 2.11\\
    Face-TTS~\cite{facetts} & \text{\sffamily X} & 82.97 & 44.37 & 2.53 & 7.44 & 8.16 & 78.24 & 39.15 & 2.48\\
    StyleDubber~\cite{styledubber} & \text{\sffamily X} & \textbf{93.79} & 18.88 & 3.73 & 5.61 & 5.69 & \textbf{86.67} & 19.58 & 3.71\\
    \midrule
    StyleSpeech*~\cite{stylespeech} & \checkmark & 91.42 & 19.63 & 3.73 & 5.77 & 5.89 & 83.73 & 22.38 &3.66\\
    V2C-Net~\cite{V2C} & \checkmark & 89.75 & 19.04 & 3.70 & 5.79 & 6.06 & 80.11 & 19.92 & 3.64\\
    \midrule
    Ours & \checkmark & 89.03 & \textbf{18.60} & \textbf{3.87} & \textbf{5.60} & \textbf{5.63} & 81.72 & \textbf{19.17} & \textbf{3.86}
    \\
    \bottomrule
    \end{tabular}
    }

  \label{grid_results}%
  \vspace{-0.4cm}
\end{table*}%
\begin{table}[!t]
  \centering
    \caption{Subjective evaluation on V2C-Animation and GRID benchmarks.}
    \vspace{-0.2cm}
  \resizebox{1.0\linewidth}{!}
  {
    \begin{tabular}{c|ccc|ccc}
    \hline
    Dataset & \multicolumn{3}{c|}{V2C-Animation} & \multicolumn{3}{c}{GRID} \\ 
    \toprule
    Methods 
    & MOS-N $\uparrow$ 
    & MOS-S $\uparrow$   
    & CMOS $\uparrow$ 
    & MOS-N $\uparrow$
    & MOS-S $\uparrow$ 
    & CMOS $\uparrow$  \\
    \midrule
    GT  & 4.52$\pm$0.13 & - & +0.19 & 4.69$\pm$0.07 & - & +0.10 \\
    GT Mel + Vocoder & 4.39$\pm$0.16 & 4.41$\pm$0.18 & +0.17 & 4.66$\pm$0.08 & 4.53$\pm$0.10 & +0.11  \\
    \midrule
    StyleSpeech*~\cite{stylespeech}  & 3.31$\pm$0.21& 3.35$\pm$0.12 & -0.25 & 3.50$\pm$0.10 & 3.58$\pm$0.11 &-0.29 \\
    Zero-shot TTS*~\cite{zeroshottts}  & 3.40$\pm$0.12& 3.47$\pm$0.18 & -0.29 & 3.58$\pm$0.21 & 3.52$\pm$0.15 &-0.26 \\
    V2C-Net~\cite{V2C}  & 3.54$\pm$0.16 & 3.51$\pm$0.18 & -0.24 & 3.62$\pm$0.06 & 3.67$\pm$0.11 &-0.22 \\
    HPMDubbing~\cite{HPMDubbing} & 3.57$\pm$0.17 & 3.54$\pm$0.12 & -0.24 & 3.77$\pm$0.20 & 3.74$\pm$0.13 &-0.19 \\
    Face-TTS~\cite{facetts}  & 3.18$\pm$0.13 & 3.24$\pm$0.16 & -0.37 & 3.39$\pm$0.21 & 3.32$\pm$0.17 &-0.32 \\
    StyleDubber~\cite{styledubber} & 3.88$\pm$0.14 & 3.90$\pm$0.14 & -0.15 & 4.02$\pm$0.11 & 4.06$\pm$0.05 &-0.09 \\
    \midrule
    StyleSpeech*~\cite{stylespeech}  & 3.31$\pm$0.21& 3.35$\pm$0.12 & -0.20 & 3.50$\pm$0.10 & 3.58$\pm$0.11 &-0.23 \\
    V2C-Net~\cite{V2C}  & 3.54$\pm$0.16 & 3.51$\pm$0.18 & -0.21 & 3.62$\pm$0.06 & 3.67$\pm$0.11 &-0.17 \\
    Speaker2Dubber~\cite{spk2dub} & 3.92$\pm$0.19 & 3.87$\pm$0.14 & -0.10 & 4.10$\pm$0.09 & 4.05$\pm$0.11 &-0.16 \\
    \midrule
     Ours  &  \textbf{4.03$\pm$0.12} & \textbf{3.99$\pm$0.08} &  \textbf{0.00}  &  \textbf{4.12$\pm$0.07}  &  \textbf{4.07$\pm$0.10}  &  \textbf{0.00}  \\
    \bottomrule
    \end{tabular}
    }
  \label{result_MOS}
  \vspace{-0.4cm}
\end{table}%

\begin{table}[!t]
\centering
\caption{Results on zero-shot test.}
\vspace{-0.2cm}
\resizebox{1.0\linewidth}{!}{
\begin{tabular}{c|c|cccc}
    \hline
    Method   & T-S & SECS(\%) $\uparrow$ & WER(\%) $\downarrow$ & UT-MOS $\uparrow$&CMOS $\uparrow$\\
    \toprule
    StyleSpeech*~\cite{stylespeech} & \text{\sffamily X} & 58.71 & 105.64 & 1.91 & -0.21\\
    Zero-shot TTS*~\cite{zeroshottts} & \text{\sffamily X} & 61.12 & 35.10 & 2.13 & -0.23\\
    V2C-Net~\cite{V2C} & \text{\sffamily X} & 41.86 & 38.72 & 1.99 & -0.19\\
    HPMDubbing~\cite{HPMDubbing} & \text{\sffamily X} & 49.31 & 106.45 & 1.35 & -0.21\\
    FaceTTS~\cite{facetts} & \text{\sffamily X} & 33.80 & 231.63 & 1.31 & -0.35\\
    StyleDubber~\cite{styledubber} & \text{\sffamily X} & 71.52 & 13.45 & 2.51 & -0.09\\
    \midrule
    StyleSpeech*~\cite{stylespeech} & \checkmark & 61.01 & 21.51 & 2.72 & -0.18\\
    V2C-Net~\cite{V2C} & \checkmark & 52.53 & 20.84 & 2.96 & -0.17\\
    Speaker2Dubber~\cite{spk2dub} & \checkmark  & 73.44 & 16.05 & 3.08 & -0.12\\
    \midrule
    Ours & \checkmark & \textbf{75.95} & \textbf{9.85} & \textbf{3.63} & \textbf{0.0}\\
    \bottomrule
\end{tabular}}
\label{zero-shot test}
\vspace{-0.6cm}
\end{table}

\subsection{Evaluation Metrics}
We employ diverse metrics to comprehensively evaluate the model's performance in the movie dubbing task:

\noindent \textbf{MCD-DTW \& MCD-DTW-SL.}
The Mel Cepstral Distortion Dynamic Time Warping (MCD-DTW) metric and its variant MCD-DTW-SL which adjusts the weights based on duration measure the distance between the generated dubbing and the ground truth following~\cite{V2C}.

\noindent \textbf{SECS.}
The speaker encoder cosine similarity (SECS) evaluates the similarity of speaker voice between the generated dubbing and the reference audio following~\cite{yourtts, SC-Glow}.

\noindent \textbf{WER.}
The Word Error Rate (WER)~\footnote{https://github.com/jitsi/jiwer} assesses the model's pronunciation accuracy by using an advanced ASR model Whisper\footnote{https://huggingface.co/openai/whisper-large}~\cite{whisper} to transcribe the generated dubbing into text and compare it with the original dubbing script.

\noindent \textbf{UT-MOS.}
UT-MOS~\cite{utmos} is a speech mean opinion score (MOS) predictor to measure the acoustic quality and naturalness of the generated dubbing following~\cite{flashspeech, naturalspeech3}.

\noindent \textbf{EMO-ACC.}
Emotion accuracy uses an emotion classifier to determine whether the emotion of the generated dubbing aligns with the ground truth labels following~\cite{styledubber}.
Note that this metric is only applicable to the V2C-Animation dataset, as the GRID dataset does not include emotion labels.

\noindent \textbf{MOS-N \& MOS-S.}
MOS-Naturalness (MOS-N) and MOS-Similarity (MOS-S) are mean option scores reported with a 95\% confidence interval based on ratings from 20 native English speakers using a scale from 1 to 5. 
Each participant is required to listen to 30 randomly selected generated dubbing and rate the dubbing according to the speech naturalness and voice similarity following~\cite{flashspeech}.

\noindent \textbf{CMOS.}
Comparative mean option score (CMOS) asks the participants to compare the dubbing generated by two models using the same input and rate them on a scale from -5 to 5 based on the criterion of matched degree between the generated dubbing with the video clip.

\begin{figure*}[t]
 \centering
  \resizebox{\linewidth}{!}{\includegraphics{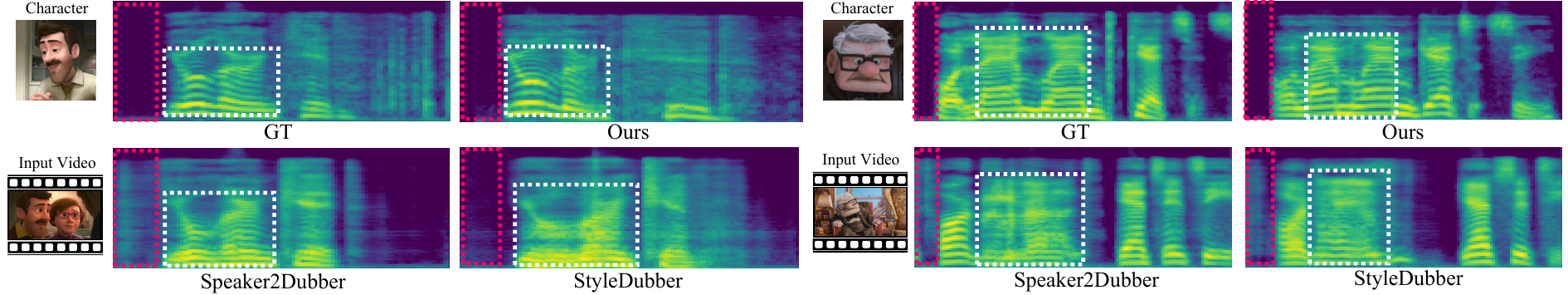}}
  \vspace{-0.6cm}
  \caption{
  The visualization of the mel-spectrograms of ground truth and synthesized dubbing by different models. 
  The red and white bounding boxes highlight regions where different models exhibit significant differences in audio quality and pronunciation details.
  }
  \label{visual}
  \vspace{-0.6cm}
\end{figure*}

\subsection{Comparison with SOTA}
\subsubsection{Results on V2C-Animation Benchmark}
As shown in Table~\ref{V2C_results}, our proposed method outperforms the current state-of-the-art models and achieves the best performance across the majority of metrics, 
Our method achieves the best pronunciation accuracy in both dubbing settings, surpasses all others including two-stage methods using the same pre-training strategy following~\cite{spk2dub}.
In terms of acoustic quality, due to the impact of inherent background noise in the movie dubbing dataset, other methods do not perform well in this aspect even with the pre-trained phoneme encoder.
However, our proposed method achieves a score improvement of 0.73-1.86 on the UT-MOS metric, bringing a significant improvement in the acoustic quality of generated dubbing.
Concurrently, our model achieves the lowest MCD-DTW and MCD-DTW-SL, indicating that the generated dubbing exhibits the smallest discrepancy and the best duration alignment with the real dubbing.
The best dubbing emotion accuracy rate which outperforms the current state-of-the-art models with a +2.58\% margin indicating the best prosody alignment performance of our method.

Although we fall short in voice similarity compared to StyleDubber~\cite{styledubber}, which learns fine-grained speaker styles, our method attains the highest score in MOS-S in subjective evaluation. 
Additionally, our proposed method also achieves the highest scores in MOS-N and CMOS, indicating the best overall dubbing quality and alignment.

\subsubsection{Results on GRID Benchmark}
Since the dubbing in the GRID benchmark is closer to traditional speech, the gap between different models is not as significant as in the V2C-Animation benchmark. 
Nevertheless, as shown in Table~\ref{grid_results}, we still achieve state-of-the-art performance in pronunciation clarity, speech quality, and discrepancy from ground truth in both dubbing settings while maintaining advanced voice similarity.
Moreover, our model also outperforms other state-of-the-art models in subjective evaluation scores on the GRID benchmark, demonstrating the effectiveness of our method across diverse dubbing benchmarks and scenarios.

\subsubsection{Results on Zero-shot Test}
In addition to dubbing benchmarks, we also conduct the zero-shot test to evaluate the generalization performance of models. 
We use audio from the GRID dataset (which is out of domain) as reference audio to dub movie clips from the V2C-Animation dataset.
Due to the absence of corresponding ground truth, we only calculate WER, UT-MOS, and SECS for objective evaluation.

As shown in Table~\ref{zero-shot test}, our proposed method surpasses the current state-of-the-art models and achieves the best performance across all metrics.
Specifically, our proposed method still ensures accurate pronunciation and the best acoustic quality of the generated dubbing when facing out-of-domain reference audio. 
In terms of voice similarity, our model achieves the best cloning effect in both objective and subjective evaluations.
The zero-shot experimental results substantiate the superiority of the generalization performance of our proposed method.

\subsection{Qualitative Analysis}
We visualize the mel-spectrograms of ground truth and dubbing generated by different models for comparison in Figure~\ref{visual}.
(More visualization results are presented in the \textit{Appendix}.)
The red bounding boxes represent regions where different models exhibit significant differences in audio quality and the white regions present the difference in pronunciation details.
Through the observation of the red bounding box, it is evident that our method exhibits less noise in the non-pronunciation intervals of dubbing compared to other models, indicating better acoustic quality.
Additionally, the clearer spectrum lines in the white bounding box and the closer resemblance to the ground truth in terms of prosody (the shape of spectrum lines) show that our method also achieves better pronunciation quality and prosody alignment performance.


\begin{table}[!t]
\centering
\caption{Results of ablation study}
\vspace{-0.4cm}
\resizebox{1.0\linewidth}{!}{
\begin{tabular}{c|ccccccc}
    \toprule
     \# & Method 
    & WER $\downarrow$ 
    & UT-MOS $\uparrow$
    & EMO-ACC $\uparrow$
    & MCD-DTW $\downarrow$ 
    & MCD-DTW-SL $\downarrow$\\
    \hline
    - & GT & 22.55 & 2.26 & 99.96 & 0.00 & 0.00 \\
    - & GT Mel+ Vocoder & 24.40 & 2.18 & 97.09 & 3.77 & 3.80 \\
    \midrule
    1 & w/o AP & 21.45 & 2.38 & 45.20 & 9.78 & 9.83\\
    2 & w/o PE & 9.84 & 3.09 & 46.12 & 9.35 & 9.38\\
    \midrule
    3 & w/o IDEA & 8.38 & 3.07 & 47.63 & 9.49 & 9.52\\
    4 & w/o PSD & 10.63 & 3.05 & 47.06 & 9.61 & 9.64\\
    5 & w/o PTBE & 8.63 & 2.60 & 46.08 & 9.84 & 9.87\\
    \midrule
    6 & Full Model &\textbf{8.04} &\textbf{3.10} & \textbf{48.93} & \textbf{9.29} & \textbf{9.32}\\
    \bottomrule
\end{tabular}}
\label{ablation study}
\vspace{-0.6cm}
\end{table}

\subsection{Ablation Studies}
To further investigate the specific effects of each module in our proposed method, we conduct ablation studies on the Dub 1.0 setting of the V2C-Animation benchmark.

\noindent \textbf{Effectiveness of Acoustic Pre-training (AP).}
As shown in Row 1 of Table~\ref{ablation study}, acoustic pre-training significantly improves the acoustic quality and pronunciation accuracy. 
It noticeably enhances the model's performance on the UT-MOS and WER. 
The clearer pronunciation and speech quality are also reflected in the improvement of dubbing accuracy indicated by MCD-DTW and MCD-DTW-SL.

\noindent \textbf{Effectiveness of Prosody Enhancement (PE).}
Row 2 of Table~\ref{ablation study} shows that the prosody enhancement strategy can reduce the difference between the generated dubbing and the ground truth, enhancing the model's performance on the MCD-DTW and MCD-DTW-SL metrics.
The results demonstrate it improves the model's acoustic modeling capability in handling complex prosody variations.

\noindent \textbf{Effectiveness of In-Domain Emotion Analysis (IDEA).}
As shown in Row 3 of Table~\ref{ablation study}, the IDEA module provides a more stable and accurate alignment between the emotion of characters and the dubbing prosody, enhancing performance on emotion accuracy. 
The better prosody alignment brought by IDEA also reduces the discrepancy between the generated dubbing with ground truth.

\noindent \textbf{Effectiveness of Prosodic Style Diffusion (PSD).}
The modeling of prosodic style features allows the model to focus on non-timbre prosodic characteristics during the prosody adapting stage, improving the overall dubbing accuracy as shown in Row 4 of Table~\ref{ablation study}.
Additionally, the disentangled design on prosodic style also enhances the model's performance in acoustic and pronunciation quality.

\noindent \textbf{Effectiveness of Prosodic Text BERT Encoder (PTBE).}
Compared to previous works that use a single text encoder output feature for both acoustic and prosodic modeling, the introduction of the Prosodic Text BERT Encoder brings disentangled text prosodic features.
This approach not only enhances the quality of text prosodic modeling but also ensures the acoustic quality of the generated dubbing.
As shown in Row 5 of Table~\ref{ablation study}, the PTBE module significantly improves the dubbing accuracy and the acoustic quality.

\vspace{-0.2cm}
\section{Conclusion}
In this work, we propose a method with a prosody-enhanced acoustic pre-training stage and an acoustic-disentangled prosody adapting stage to generate high-quality movie dubbing with aligned prosody.
The first stage equips the model with the acoustic modeling capability to handle complex dubbing prosody.
The second stage employs a disentangled prosody adapting framework to bridge character performances with dubbing prosody while maintaining the acoustic quality.
Extensive experimental results demonstrate the effectiveness of our proposed method.

\section*{Acknowledgement}
 \begin{sloppypar}
This work was supported by National Natural Science Foundation of China: 62322211, 62336008, U21B2024,
``Pioneer'' and ``Leading Goose'' R\&D Program of Zhejiang Province (2024C01023, 2023C01046, 2023C01038), 
Key Laboratory of Intelligent Processing Technology for Digital Music (Zhejiang Conservatory of Music), Ministry of Culture and Tourism (2023DMKLB004).
Yuankai Qi and Anton van den Hengel are not supported by the aforementioned fundings.
\end{sloppypar}

{
    \small
    \bibliographystyle{ieeenat_fullname}
    \bibliography{main}
}

\end{document}